\documentclass{llncs}
\usepackage{makeidx}  

\usepackage{epsfig}
\usepackage{ulem}
\usepackage{graphicx}

\usepackage{amssymb,amsfonts}
\usepackage{amsmath}
\usepackage{algorithm}
\usepackage{algpseudocode}
\usepackage{listings}
\usepackage[utf8]{inputenc}
\usepackage[USenglish]{babel} 
\begin{document}

\title{Virtualization technology for distributed time sensitive domains}

\author{ 
  Carlos Antonio Perea-G\'{o}mez 
}
\institute{Universidad Carlos III de Madrid\\Av. de la universidad 30, 28911 Leganes, Spain\\ 
\email{carlosantonio.perea@alumnos.uc3m.es}
}

\maketitle              

\begin{abstract}

This paper reports on the state of the art of virtualization technology for both general purpose domains as well as real-time domains.
There exits no entirely instantaneous data transmission/transfer. There always exist a delay while transmitting data, either in the processing or in the medium itself. However most systems are designed to function appropriately with a delay tolerance. This delay, inevitably, is affected when operating with an extra layer, the virtualization.\\ 
For real time systems it is crucial to know the temporal limits in order not to surpass them. Introducing virtualization in the real-time domain therefore requires deeper analysis by making use of techniques that will offer results with deterministic execution times. The study of time in systems and its behaviour under various possible circumstances is hence a key for properly assessing this technology applied to both domains, especially the real-time domain.\\

\end{abstract}


\section{Introduction}
Technology must keep up with the needs of the new innovative optimized systems. An increasing tendency in the usage of pure electronically driven systems for handling complex multifunctional systems, such as an aircraft or an automobile, can be observed. For instance traditional steering or throttle, in the automotive sector, has been modified with electronically driven systems. In aircraft, fly-by-wire technology, or Intelligent Flight Control System by NASA \cite{ifcs}, have been researched and developed to optimize the traditional manual flight controls.\\
Stability, reliability and stand-alone ability are some of the required features  by modern systems. These systems are the result of combining complex hardware components with multiple computational units which are in charge of the arithmetic calculations, logic, I/O and control operations.\\
It is the case that a full multifunctional system must be driven by a large amount of different systems, or sub-systems, in addition to their redundant backups to reduce their points of failure. This large amount  has a significant negative impact on the cost effectiveness, weight and required physical space.\\
With the design and development of such multifunctional systems comes the necessity of combining multiple physical computational units. As a consequence the degree of difficulty when optimising some basic parameters such as performance, functionality and cost effectiveness rises.  Stacking up the various units and interconnecting them might be at first glance the most straight forwarded and simplex solution. This way the required features such as performance are guaranteed. however a different approach has to be taken if an optimum trade-off of the parameters ought to be achieved.\\ \newline
A possible workaround is to make use of virtual environments. By way of using a single physical unit instance to hold and populate various virtual instances in charge of running the respective sub-systems. These virtual instances will hence share the physical resources such as processing units and memory. A correct configuration and scaling of such virtual environment may allow to achieve the required performance specifications while decreasing cost, saving weight and reducing space.\\
Nevertheless this walk-around rises yet a further difficulty. Keeping up the performance specifications of the different virtual instances at all moment in time may not be as uncomplicated as it may seem initially. There are many factors that must be examined in depth before asserting whether a system is adequate for a virtual environment. Regardless of the sufficiency of the more obvious aspects in a virtual environment (virtual processing units, memory, shared storage among others) is the connectivity between the different virtual instances or sub-systems.\\ \newline
Inside an automobile or an aircraft complex data cross reference operations are made to calculate millions of instructions. This data is sent across the interconnected sub-systems. These operations require perfect connectivity patterns. Furthermore time is a critical factor, milliseconds, even microseconds may change the course of outcome in a specific situation. The electronic stability control (ESP) may be taken as an example. Continuous monitoring and comparison of two data samples, the intended and the actual moving direction, helps keep a automobile, under any circumstance, on track. In the event of ice on the road, the actual moving direction is affected, the system detects this discrepancy  when comparing it with the intended moving direction and reacts adequately correcting the trajectory almost instantaneously. This way the automobile is kept safe on track. An absolute failure of the data communication, or simply a delay, may have indeed fatal consequences.\\ \newline
The previous example can be used analogously with other systems and helps to understand how important the time factor is.\\
Here lies the crucial aspect of virtualization. Is data transmission performance using virtual environments sufficient? Does it match the reliability of traditional communication? Does virtualization add a noticeable delay to the transmissions? This paper examines virtual environments and their communication capabilities in order to be able to be able to draw conclusions to these questions. It is structured as follows. 
Section \ref{sec:Overview} provides a survey on selected virtualization technologies describing particular aspects of virtualization software for real-time domains. Section \ref{sec:Development} presents experimental set up.
Section \ref{sec:Results} provides experimental results and draws the conclusions. Lastly Section \ref{sec:Background} describes selected related work and Section.

\section{Virtualized distributed systems}
\label{sec:Overview}

\subsection{Virtualization technologies}

As briefly mentioned in the introduction, virtualization is the  act of creating a virtual version of the systems. One physical platform  allows to hold multiple virtual machines which are run by software packages such as hypervisors and emulators. Various types of virtualization techniques have been developed. The two techniques more often used are the so called paravirtualization and hardware virtual machine virtualization. Do note that there are variations and combinations of these techniques which result in further techniques.
\begin{itemize}

\item  \textbf{Paravirtualization (PV)}: efficient and lightweight technique that does not require to emulate the full set of hardware and firmware services. Guest operating systems are aware of the hypervisor and run more efficiently without the emulation nor the virtualization of HW. There is no explicit virtualization support, however PV-Kernel and PV-drivers are essential.
\item  \textbf{Hardware Virtual Machine (HVM)}: also known as full virtualization, makes use of virtualization extensions from the host CPU to achieve guest virtualization. These extensions are used for performance boost purposes. The Hardware components, adapters and BIOS are emulated while kernel support is not required.
\end{itemize}
\subsection{Virtualization platforms}
Additional to these techniques there are over 50 different platforms that allows to undergo virtualization in the systems. the scope of this paper will focus on \textit{KVM},\textit{QEMU} and \textit{Xen} for general purpose domains, and \textit{XtratuM} for real-time domains.\\
\newline
\textit{\textbf{QEMU}} is an emulator and virtualization machine that allows you to run a complete operating system as just another task on your desktop. It can be very useful for trying out different operating systems, testing software, and running applications that won't run on your desktops native platform.\\
QEMU runs on x86 systems running Linux, Microsoft Windows, and some UNIX platforms, and can host target systems from a range of different microprocessors.[5]\\ \newline
\textit{\textbf{KVM}}, Kernel-based Virtual Machine, is a full virtualization solution for Linux on x86 hardware containing virtualization extensions, Intel VT or AMD-V. It consists of a loadable kernel module, kvm.ko, which provides the core virtualization infrastructure and a processor specific module, kvm-intel.ko or kvm-amd.ko. Using KVM, multiple virtual machines running unmodified Linux or Windows images can run. Each virtual machine has private virtualized hardware, network cards, disks, graphics adapters an so on.[4]\\
It is important to note that KVM is a fork of Qemu executable. A system can run Qemu by itself and it will handle all the virtual machine resources as well as all the virtual HW and CPU. The only drawback is the extreme slow communication between the host and guests CPU's. This aspect can be optimized using KVM on top, which only focuses on the acceleration of the CPU.\\
If we define the hypervisor or Virtual Machine Monitor (VMM) as a system that creates and runs VM then we can affirm that QEMU is a fully standalone hypervisor whereas KVM is just an accelerator that uses processor extensions.\cite{kvm}\\
\newline
\textit{\textbf{Xen}}, now called Xen Project is a native type 1 hypervisor which makes it possible to run various instances, or rather different operating systems, on a single physical machine. It makes use of the paravirtualization technique. Server and desktop virtualizations as well as IaaS applications are the most common direct application of this hypervisor.\cite{xen}\\
\begin{figure}[h]
\begin{center}
\includegraphics[width=0.6\textwidth]{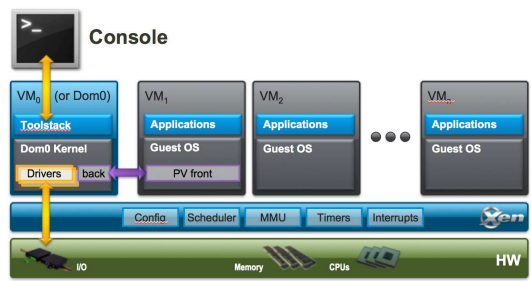} \\
\end{center}
\caption{Xen Architecture \cite{xen}}
\label{fig:xen_arh}
\end{figure}
\newline
Due to the microkernel design Xen only generates a small memory footprint, around 1MB, and restricts a limited interface to the guests. This enforces the robustness and security of the hypervisor. Additionally it provides driver isolation which guarantees that a driver's fault within a VM does not affect the rest of the system.\\ 
The hypervisor runs directly on the HW and is responsible for managing the memory, CPU and diverse I/O interrupts. The domain 0 is a special domain that apart from containing the drivers will also control and manage creation, destruction and configuration of further virtual machines of the system via means of the toolstack.\cite{xen}\\
\newline
\textit{\textbf{Xtratum}} is a real time hypervisor, also of type 1, that provides a framework to run several real-time executives in a robust partitioned environment. Designed as a nanokernel, it virtualizes the essential hardware components for the execution of several concurrent operating systems, being at least one a real time Os. Additionally in order to reduce the design complexity and increase the reliability, its nanokernel was designed as a monolithic and nonpreembale. It is important to note its Independence from Linux and its ability to be bootable.\cite{xtratum}\cite{xm1}\\
\newline Xtratum meets safety critical real-time requirements and hence is used to build partitioned systems. A partition is defined analogous to a virtual machine or instance, as an execution environment managed by the hypervisor. Each partition can make use of the full physical resources at given times, and can support multiple processes implemented by the guest OSes. In order for these partitions to be executed on the hypervisor, they need to be virtualized. The applications, or the execution code must be written the corresponding partition. Additionally, XtratuM takes control of the system at boot time and initializes the hardware prior to executing the partition code.\\
A further relevant feature of XtratuM is the strong temporal  and spatial isolation. Temporal isolation is achieved via a fixed cyclic scheduler which makes it impossible for an application to run in parallel with one-another. Strong spatial isolation prevents the sharing of memory and enforces each partition to only access its own allocated memory. Additionally it also provides a strong robust communication mechanism.\cite{xm6} \\
This robust communication mechanism between partitions is a port-based communication. XtratuM implements the channel between at least two access points or ports. Hence  the channel is only accessible by the different partitions containing those ports. It is the hypervisor responsibility to encapsulate and transport the data through the channels. There are two different modes in which communication can occur: sampling and queuing.\\
\newline
Xtratum also provides  with a Fault Management Model. Fault shall include events of a system trap including HW and SW interrupts including processor interrupts, or an event triggered by the hypervisor itself. Faults are at first detected and handled by the hypervisor, then propagated to the corresponding partitions. Furthermore a Health Monitor, part of the hypervisor, is in charge of detecting and reacting to anomalous events or states.\cite{xm6}\\
As a result of the spatial isolation of the different partitions subsystems will not be affected by a faulty partitions.\\
XtratuM uses a cyclic scheduler to run the various partitions, therefore the system consist of states and transitions. During the initialization the partitions find themselves in the \textit{Boot} state. There the initialization of the virtual machine and the standard execution environment including communication, I/O, interrupts and more services occur. After booting the partition changes to \textit{Normal} state. At this state the code is ready to be executed whenever demanded by the hypervisor. The transition between the execution of the different partitions happens automatically according to the fixed time schedule predefined during the initial set up of the environment.\cite{xm4} \\
\newline
The simplest scenario for developing XtratuM is composed of two different actors, a partition developer (PD) and an integrator (I). The interaction of these two will allow the establishment of a correct working environment for XtratuM.  \\
The first step is for the PD to define the required resources and inform this to the integrator. With this information the integrator configures the XtratuM source code, adapting it to the requirements and using libraries and tools builds the hypervisor binary, \\
Additionally the integrator then allocates the available system resources to the partitions. This is creates the \textit{XM CF} configuration file, where memory areas, scheduling plans, port and channel creation among other resources are detailed described for further processing.\cite{xm5}\\ 
These resulting binaries are unique and are then sent back to the PDs for the development and building of the execution environment. Once the environment is built, the PD can move on, and creates the partition application. Lastly the integrator will pack the gathered binaries and image together with the resident software. This will allow the partition to be deployed and run.\\
This work-flow has been described in a very simplistic manner. Each step described, includes multiple sub-steps and the use of the extensive XtratuM tool's library.\cite{xm5}\\

\subsection{Communication software or middleware}

\begin{figure}[H]
\begin{center}
\includegraphics[height=6cm]{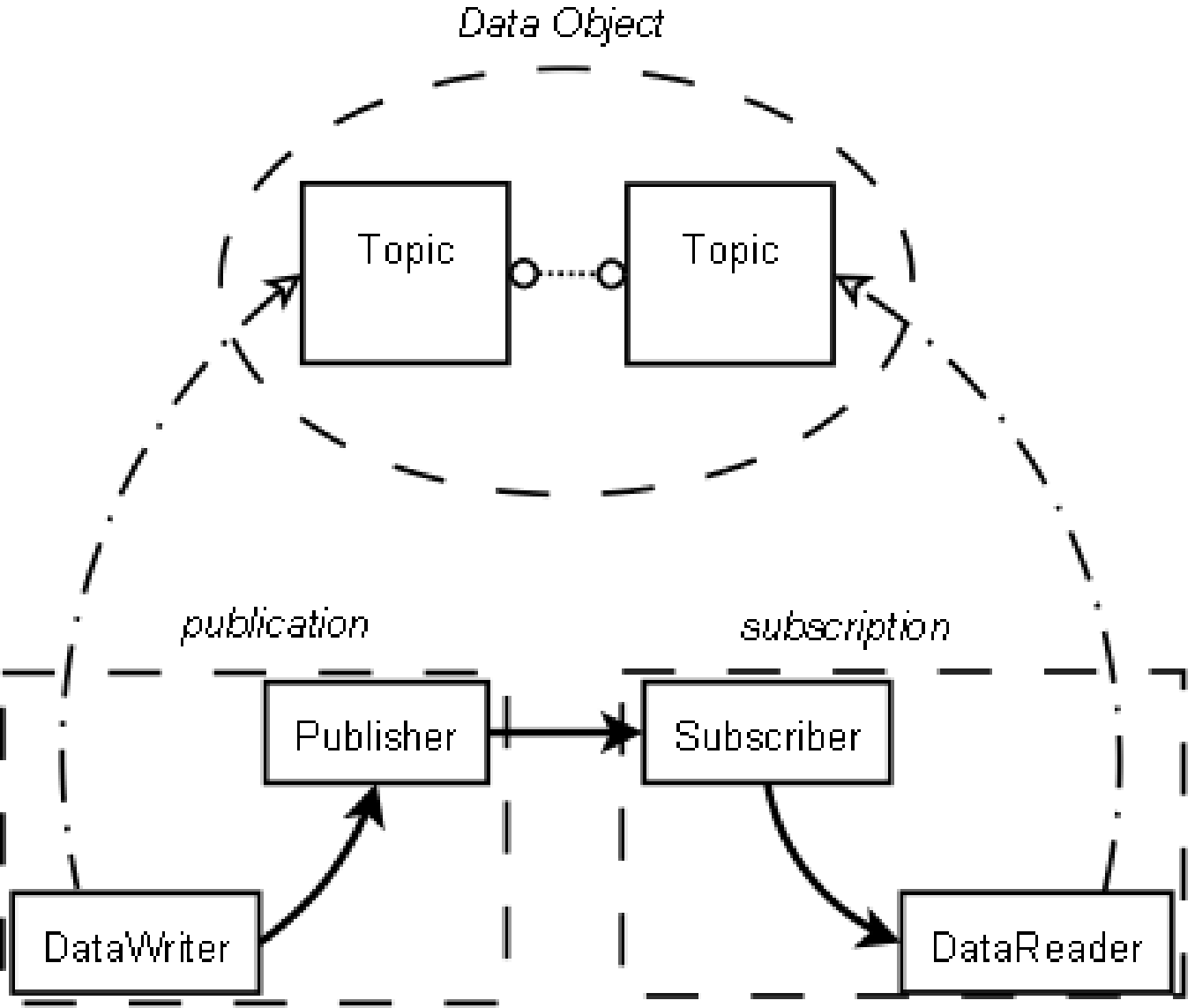} 
\end{center}
\caption{DDS\cite{dds}}
\label{fig:dds-arch}
\end{figure}

The middleware can be described in its wide sense as software glue. It lies between the OSes and applications and enables multiple components of a system to communicate and share data.\\
\newline
\textit{\textbf{Internet Communication Engine}},Ice, is an object oriented middleware platform. It provides tools, APIs and library support to be used in heterogeneous environments, being unnecessary that both emitter and receiver are written in the same programming language. It can run on different OSes and machine architectures and allows  communicating using various networking technologies.\cite{ice} \\
Each Ice object contains an interface with a certain number of operations. These operations, as well as interfaces and data types are exchanged between both ends, client and server, using the Slice language.This language is purely declarative and no statement for execution can be written. The translation of Slice into the different programming languages is supported for C++, C\#, Java, JavaScript, Python, Objective-C, PHP and Ruby.\\ 
\noindent The IceStorm Service simplifies substantially the implementation of data transfer between multiple clients. It presents a publisher-subscriber solution. It acts as a mediator between the publisher and subscribers respectively. This way the clients are only responsible for sending or receiving data to or from the server. The server will then be in charge of the distribution of the data which allows the clients to disregard any further action rather than sending one set of data to the server.\cite{ice}\\
\newline
\textit{\textbf{Data Distribution Service}}, DDS, is a data communication standard managed by Object Management Group (OMG) that provides scalable, low-latency, high performance and interoperable data exchange for distributed applications. It is suitable for real time and near real time systems.\cite{dds} \\
The DDS Standard includes a well defined API that allows the creation of portable code. It references the Real Time Publish Subscribe (RTPS) Wire Protocol standard which defines the wire protocol for DDS communications. Generally speaking DDS is a p2p communication model requiring no gateway, server nor daemons that have to be running nor configured. \\
\textit{OpenDDS} is an open-source version supporting the capability defined in the DDS 1.2 Specification and version 2.2 of The RTPS DDS Interoperability Wire Protocol Specification. A very basic conceptual view of the mentioned Publish Subscribe Architecture can be seen in figure \ref{fig:dds-arch}.\\

\noindent Basic components are: Topic, Publisher, Subscriber, DataWriter and DataReader. Topics contain information about a single data type, the distribution and availability. Publication or a Subscription can have many associated Subscriptions or Publications receptively. It is important to note that an application can be either a Publisher a Subscriber or eventually both. Hence the data flow can be easily understood by following the arrows in the conceptual diagram. The association of a publication and a subscription exists only when compatible Topics meet at both writer and reader side.. In this case the Publication and Subscription become associated and data can be exchanged.\\

\section{Developing a distributed partitioned/virtualized system}
\label{sec:Development}
This section includes a high level description of the procedure taken to develop and implement a distributed environment. As mentioned, the scope of this paper includes both real-time and general purpose domains, therefore two different setups will be implemented. Also as briefly explained in Section \ref{sec:Overview}, by definition real-time systems are described as being partitioned rather than virtualized.\\
For a common understanding a distributed virtual system is a system combined of multiple components across a network which work together. In this case these components will be virtual rather than physical which will be interconnected. A distributed system can be hence translated into a virtual environment.\\
\newline
The requirements will therefore be  a physical computer unit with the correspondent compatible hardware for the virtualization software. Additionally as the focus is to deep-dive into the transmission performance aspects, middleware technology will be enforced. This will ensure communication scenarios for analysis purposes. \\
\newline
\textit{\textbf{Setup A}} will server as the distributed environment for the analysis of general purpose domain. For the implementation Xen 4.4 hypervisor is used to run various domains using the Ubuntu Xenia Xersus distribution. The last piece of the puzzle, the middleware, consists of Ice's service Icestorm which include the components of the subscriber, publisher and server. Version 3.6 is used \cite{Ice36}\\
The various described software components require different overlapping minimum hardware specifications. In order to ensure that everything will work the most restrictive minimum specs of all different aspects must be chosen. Xen, for example, imposes the more restrictive minimum requirement of 1 GB RAM, regardless of whether Ubuntu will work just fine with half of that.\\
The decision for this setup lies in the fact that Xen allows to perform a clean paravirtualization.\\
Additionally Icestorm offers the necessity to have at least three different components that will form a clean distributed virtual system with coherent roles in each of the component members. \\
\begin{table}[]
\centering
\caption{Unit A}
\label{tab:setupa}
\begin{tabular}{ll}
\hline
\textbf{Chipset} & \textit{Intel Core 2 Duo P8700} \\ \hline
\multicolumn{1}{l|}{\textbf{Performance}} & \textit{2 x 2.53 GHz} \\ \hline
\textbf{RAM} & \textit{1536 MB} \\ \hline
\end{tabular}
\end{table}
For the multiple domain configuration and installation of the Ubuntu distributions, configuration files have to be created appropriately. These config files include all hardware parameters such as memory allocation, network configuration and also the paths to the distribution installation package.\\
On command, Xen parses the configuration file and installs the corresponding domain as described on the configuration file.\\
The different domains will hold the different components, publisher, subscriber and server. This latter one being in charge of the correct distributing of the data.\\
The intercommunication among the various domains implemented by means of the the Xen Toolstack. Brdige-utils allows to implement a virtual switch and virtual network interfaces per guest domain, and so allow communication.\\
With this setup the pertinent performance tests are ready to be executed.\\
\newline
\textit{\textbf{Setup B}} includes the implementation of a portioned system on the Xtratum hypervisor, version \textit{xmvm-x86-2.4}. Similar to the first setup hardware specifications have to be met. In this case the thought of concurrent computation is no longer valid, hence a different mind-set must be triggered to better understand this system.\\
The hypervisor can be install on a bare machine, and via the configuration fies, in this case an xml file, the partition cyclic schedule can be configured. A single cycle, milliseconds, in the configuration is represented with the total duration of the cycle and the duration of each of the partitions. Moreover in this file the correspondent ports, channels and resource allocation has to be detailed. After a correct implementation of this xml file, Xtratum is in charge of bringing up the various partitions and keeping the cycles.\\
\begin{figure}[H]
\begin{center}
\includegraphics[width=0.6\textwidth]{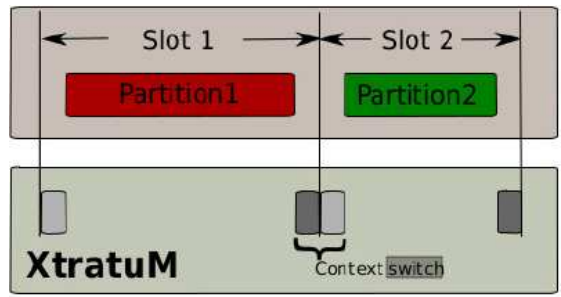} 
\end{center}
\caption{Cycle Description \cite{xm5}}
\label{fig:xt_cycle}
\end{figure}
\noindent For simplicity purposes, and in order to obtain clean deterministic results, just simple applications will be installed on the partitions. This means that no real Os will be installed, but rather the simple applications of publisher and subscriber. This two applications will be communicating via a defined channel using the dedicated virtual ports. Hence port based communication. One of the applications will send data to the other.\\
Internally the communication procedure implies that when the publisher is active, the applications writes data to a dedicated buffer. When the subscriber application is active, it will read the data from the buffer. It is important to remember that all parameters, ports, channel as well as the size of such buffer has to be described in the xml file.\\
In other words this second setup can be seen as a more traditional system, with the singularity of not running parallel components, but still communicating by simply writing on a buffer and reading from it.\\

\section{Experiments and Results}
\label{sec:Results}
With the implemented setups of the two different scenarios, over 100 repetitions of performance tests were carried out to make a detailed analysis of the impact on the transmission times. In order to set ground zero, multiple data samples of 1 Byte, 1 Mbyte and 6 Mbytes were transmitted and its transmission times were recorded.\\
\newline
For \textit{\textbf{Setup A}}, as the general purpose domain is under analysis the absolute transmission times do not offer any interesting information. In order to obtain meaningful results the distributed system was set under full CPU load, and around 75\% of memory usage. By means of the following formula, comparing results of idle CPU transmission and loaded CPU, results in the transmission delay can be obtained\\
$$ Tx Delay = Stressed Tx Time - Relaxed Tx Time$$
An overall average value of just under 5 ms in transmission delay was obtained. With this information conclusions can be drawn on how well distributed virtual systems perform, and what its impact in the transmission times are.\\
\newline
As for \textbf{\textit{Setup B}} a similar performance test was carried out. The only difference is the lack of middleware and the complexity of having to adjust the channel buffer accordingly to the amount of data wanting to be sent. No CPU load increase was possible as just simple bare applications were running on the different partitions. Nevertheless combinations of different cyclic schedule were made to observe its behaviour.\\
As the real-time domain is under analysis, the actual absolute delay of the transmission is of high importance. For interpreting the results obtained, the understanding of the cyclic partitioned functionality is crucial. As there is a fixed duration predefined for each partition to be executed each cycle, the transmission time should is indirectly set by design, and corresponds to the transition time from one partition to the other partition. In other words, for the system to maintain the real-time aspect, the transmission of data should not exceed the actual time set between the execution of these two partitions.\\
The results obtained indicate a total transmission delay of 2.1\% compared to the transition time between partitions. In absolute values this corresponds to just a few $\mu s$ which are indeed negligible. The results obtained confirms the time criticality of this hypervisor, and indicates the possibility to implement such for real-time systems as the dala is far from being significant. However a further important finding indicates anomalies when the actual execution time of the application within the partition exceeds the the predefined duration of the partition itself. \\


\section{Background}
\label{sec:Background}

The presented paper is framed in the context of the research performed by the group on time-sensitive distributed  large scale systems of Universidad Carlos III de Madrid, and precisely on middleware design, resource management, and analysis of virtualization technologies for real time distributed systems and cyber-physical systems. More precisely, this work is based on the challenges, open problems, and solutions identified in \cite{RTCloud,JSASI2015,JSASI2017,FGCSSI2017} for middleware technologies both over bare machines and virtualized envirornments for predictable cloud computing, and in \cite{Omacy} for middleware technologies in cyber-physical systems. \\
\newline
Different middleware technologies have served as basis for design improvements. For example, Ice (Internet Communication Engine) middleware was improved in various contributions such as \cite{Sac2016} in which different parameters were adjusted to improve performance. Moreover, in \cite{Isorc2016} a centralized architecture for real time distributed systems was designed to support a variable (and larger) number of clients. Additional proposals were made for online verification of cyber-physical systems connected as in \cite{Compsac2014}, \cite{Indin2014} and \cite{Hase2016} as well as middleware with augmented logic for supporting changes in the dynamic structure of real time distributed systems based on services 
\cite{iLandTII}.\\
\newline
The current report summarizes a larger work enclosed within the efficient management of the execution platform resources, not only in the participating nodes but also in the interaction and communication between the nodes by means of the improved middleware technology. Likewise, it is enclosed within the dynamic management which requires augmented logic \cite{Vision2011}.\\
\newline
With the management of the resources in the participating nodes, a better exploitation of the computational resources (CPU, memory, energy, etc.) is obtained using priorities and resource managers as in \cite{HolaQoS}, \cite{PhD}, \cite{ARES98}, \cite{ModeChange}, \cite{dualBand}, \cite{Clara}, \cite{Cano2013}, 
or \cite{mobileOS}. Analysis of the kernel functions has also been performed for the kernel itself \cite{AdaEurope2004} and for the networked storage \cite{SPE2006}. Integration of a tight resource control has been recently performed to make the communication middleware aware of the multicore nature of the underlying processors \cite{Sac2017}.\\
\newline
In the management of the communication between the nodes based on middleware, for a system with time requirements, or in general for real time systems it is necessary to analyze the reliability of the platform, its performance and temporal stability. In this context the following projects have been developed. 
In \cite{ACMSigbediLand} an eHealth application was developed supporting reconfiguration triggered by sensors; this work has been recently enhanced with a component model \cite{Compsac2017} for the Integrated Clinical Environment. The temporal cost of the communication of sensors and embedded computers such as Rasperry Pi has been analyzed in \cite{gpcRaspi} and the influence of the middleware design in the communication costs has been analyzed in 
\cite{gpcUMA,DDSPartitioned}
A bridge for adapting middleware to different communication paradigms is presented in \cite{AdaEurope2011}. And lastly, augmented interconnection between iLand and the distributed annex Ada (Ada DSA) is presented in \cite{AdaEurope2012}.\\

\section{Conclusions}

This report has presented a summary of a larger work on analyzing the temporal behavior of distributed virtualized systems making use of middleware over real-time virtualization technology. The paper has presented an overview of virtualization technology for efficient, low footprint, and high performance systems, as well as middleware technology that is used to enable distributed communication. The paper has also reported on some summarized results on an actual setting for a distributed virtualized/partitioned system.


\begin{thebibliography}{5}

\bibitem{ifcs} 
NASA Dryden Fact Sheet - Intelligent Flight Control System\\
\textit{http://www.nasa.gov/centers/armstrong/news/FactSheets/FS-076-DFRC.html}\\
NASA, Gibbs, Yvonne

\bibitem{kvm} 
Linux-KVM\\ 
\texttt{http://www.linux-kvm.org/page/MainPage}

\bibitem{xen} 
Xen Project Wiki. Support and Documentation Resources\\ 
\texttt{http://wiki.xen.org/wiki/XenOverview}

\bibitem{xtratum} 
XtratuM Hypervisor\\ 
\texttt{http://www.xtratum.org/}

\bibitem{xm1} 
M. Masmano, I. Ripoll, A. Crespo, J.J. Metge, and P. Arberet.  Xtratum:  An opensource hypervisor for tsp embedded systems in aerospace.  In DASIA 2009.  DAta SystemsIn Aerospace., May.  Istanbul 2009.

\bibitem{xm4}
FENTISS /UPVLC. XtratuM Hypervisor for INTEL x86. Volume 3: User Manual. September 2012.

\bibitem{xm5}
FENTISS /UPVLC. XtratuM Hypervisor for INTEL x86. Volume 3: Reference Manual. September 2012.

\bibitem{xm6}
M. Masmano, J. Coronel, fentISS, Valencia, Spain, P. Balbastre, A. Crespo, J. Simo, S. Peiro
Universitat Politecncia de Valencia, Spain "XtratuM hypervisor for mixed-criticality systems"

\bibitem{dds}
Open DDS Overview\\
\texttt{http://www.opendds.org/about.html}

\bibitem{ice} 
Zeroc Inc. Ice 3.6.1 Documentation Manual\\ 
\texttt{https://download.zeroc.com/Ice/3.6/Ice-3.6.1.pdf}

\bibitem{Ice36}
ZeroC Inc.
\textit{The Internet Communications Engine}. \texttt{https://zeroc.com/downloads/ice/3.6/} (on-line).
2016.

\bibitem{RTCloud}
M. Garc\'{i}a-Valls, T. Cucinotta, C. Lu.
\textit{Challenges in real-time virtualization and predictable cloud computing}.
Journal of Systems Architecture, vol. 60(9), pp. 726--740. 2014.

\bibitem{iLandTII}
M. Garc\'{i}a-Valls, L. Fern\'{a}ndez Villar, I. Rodr\'{i}guez L\'{o}pez.
\textit{iLAND: An enhanced middleware for real-time reconfiguration of service oriented 
distributed real-time systems}. Transactions on Industrial Informatics, vol. 9(1), pp. 228-236. 2013.

\bibitem{HolaQoS}
M. Garc\'{i}a-Valls, A. Alonso, J. Ruiz, A. Groba. 
\textit{An architecture for a quality of service resource manager middleware 
for flexible multimedia embedded systems}
Proc. $3^{rd}$ Int'l Conference on Software Engineering and Middleware (SEM). 
LNCS, vol. 2596, pp. 36--55. 2003.

\bibitem{Omacy}
M. Garc\'{i}a Valls, R. Baldoni.
\textit{Adaptive middleware design for CPS: Considerations on the OS, 
resource managers, and the network run-time}.
Proc. $14^{th}$ Workshop on Adaptive and Reflective Middleware (ARM).
Co-located to ACM ACM/IFIP/USENIX Middleware. 
Vancouver, Canada. December  2015. 

\bibitem{Cano2013}
J. Cano, M. Garc\'{i}a-Valls.
\textit{Scheduling component replacement for timely execution in dynamic systems}.
Software: Practice and Experience, vol. 44(8), pp. 889-910.
January 2013.

\bibitem{dualBand}
M. Garc\'{i}a-Valls, A. Alonso, J. A. de la Puente. 
\textit{A dual priority assignment mechanism for dynamic QoS resource management}. 
Future Generation Computer Systems, vol. 28(6), pp.902-911. June 2012.

\bibitem{Clara}
C. M. Otero P\'{e}rez, L. Steffens, P. van der Stok, S. van Loo, 
A. Alonso, J. Ru\'{i}z, R. J. Bril, M. Garc\'{i}a Valls.
\textit{QoS-Based Resource Management for Ambient Intelligence}.
In: \textit{Ambient Intelligence: Impact on Embedded Sytem Design}, pp. 159--182.
Kluwer Academic Publishers. 2003.

\bibitem{PhD}
M. Garc\'{i}a-Valls.
\textit{Calidad de servicio en sistemas multimedia empotrados mediante gesti\'{o}n 
din\'{a}mica  de recursos}. Tesis doctoral. Universidad Polit\'{e}cnica de Madrid. (2001)

\bibitem{ARES98}
A. Alonso, M. Garc\'{i}a-Valls, J. A. de la Puente.
\textit{Assessment of Timing Properties of Family Products}. 
ESPRIT ARES Workshop 1998, LNCS vol. 1429, pp. 161--169. Las Palmas, Spain. 1998.

\bibitem{ModeChange}
M. Garc\'{i}a-Valls, A. Alonso, J.A. de la Puente.
\textit{Mode change protocols for predictable contract-based resource management in 
embedded multimedia systems}. 
In Proc. of IEEE Int'l Conference on Embedded Software and Systems (ICESS), pp. 221-230. 
May 2009.

\bibitem{Isorc2016}
M. Garc\'{i}a-Valls.
\textit{A proposal for cost-effective server usage in CPS in the presence of dynamic 
client requests}. 
Proc. of $19^{th}$ IEEE International Symposium on Real-time Distributed Computing (ISORC).
York, UK. May 2016.

\bibitem{Hase2016}
M. M. Bersani, M. Garc\'{i}a-Valls.
\textit{The cost of formal verification in adaptive CPS. An example of a virtualized server node. }
Proc. of $17^{th}$ IEEE High Assurance Systems Engineering Symposium (HASE). 2016.

\bibitem{SPE2006}
P. T. Breuer, M. Garc\'{i}a-Valls.
\textit{Raiding the Noosphere: the open development of networked RAID support for the Linux kernel.}
Software Practice and Experience, vol. 36(4), pp. 365--395. Wiley. 2006.

\bibitem{AdaEurope2004}
P. T. Breuer, M. Garc\'{i}a-Valls.
\textit{Static Deadlock Detection in the Linux Kernel.}
Ada Europe. LNCS vol. 3063, pp. 52--64. Palma de Mallorca, Spain. 2004.

%
%
%
%
%



\bibitem{AdaEurope2011}
I. Rodr\'{i}guez-L\'{o}pez, M. Garc\'{i}a-Valls.
\textit{Architecting a Common Bridge Abstraction over Different Middleware Paradigms}. 
Ada-Europe 2011, pp. 132-146. Edimburgh, UK. June 2011.


\bibitem{AdaEurope2012}
M. Garc\'{i}a-Valls, F. Ib\'{a}nez-V\'{a}zquez.
\textit{Integrating Middleware for Timely Reconfiguration of Distributed Soft Real-Time 
Systems with Ada DSA}. 
Ada-Europe 2012, pp. 35-48. Stockholm, Sweden. July 2012.




%


\bibitem{mobileOS}
M. Garc\'{i}a-Valls, A. Crespo, J. Vila.
\textit{Resource management for mobile operating systems based on the active object model}.
International Journal of Computer Systems Science \& Engineering, vol. 28(4), 195--205.
2013.


\bibitem{Indin2014}
M. Garc\'{i}a-Valls, D. Perez-Palacin, R. Mirandola.
\textit{Extending the verification capabilities of middleware for reliable distributed self-adaptive systems}. 
Proc. of $12^{th}$ IEEE International Conference on Industrial Informatics (INDIN).
Porto Alegre, Brazil. July 2014.  

\bibitem{Vision2011}
M. Garc\'{i}a-Valls, F G\'{o}mez-Molinero.
\textit{Real-time reconfiguration in complex embedded systems: A vision and its 
reality}.
Proc. of $9^{th}$ IEEE International Conference on Industrial Informatics (INDIN).
Lisbon, Portugal. July 2011.

\bibitem{Compsac2014}
M. Garc\'{i}a-Valls, D. Perez-Palacin, R. Mirandola.
\textit{Time sensitive adaptation in CPS through run-time configuration generation and verification. }
Proc. of  $38^{th}$ IEEE Annual Computer Software and Applications Conference (COMPSAC), pp. 332--337. Vasteras, Sweden. July 2014.

\bibitem{Sac2016}
M. Garc\'{i}a-Valls, C. Calva-Urrego, J. A. de la Puente, A. Alonso. 
\textit{Adjusting middleware knobs to assess scalability limits of distributed cyber-physical systems.} 
Computer Standards \& Interfaces, vol. 51, pp. 95--103. March 2017.

\bibitem{Sac2017}
M. Garc\'{i}a-Valls, C. Calva-Urrego, J. A. de la Puente, A. Alonso. 
\textit{Improving service time with a multicore aware middleware}. 
$32^{nd}$ ACM Annual Symposium on Applied Computing. Marrakech, Morocco. April 2017. (To appear)

\bibitem{Compsac2017}
I. E. Touahria, M. Garc\'{i}a-Valls, A. Khababa.
\textit{An ICE compliant component model for medical systems development. }
Proc. of  $41^{st}$ IEEE Annual Computer Software and Applications Conference (COMPSAC). 2017. (To appear)

\bibitem{JSASI2017}
M. Garc\'{i}a-Valls,  A. Casimiro, H. P. Reiser.
\textit{A few open problems and solutions for software technologies for dependable distributed systems.} 
Journal of Systems Architecture, vol. 73, pp. 95--103. January 2017.

\bibitem{JSASI2015}
M. Garc\'{i}a-Valls,  T. Cucinotta.
\textit{Real-time and distributed computing in emerging applications. Foreword by the general chairs of Reaction 2012.} 
Journal of Systems Architecture, vol. 61, pp. 267--268. May 2015.

\bibitem{FGCSSI2017}
M. Garc\'{i}a-Valls,  P. Bellavista, A. Gokhale.
\textit{Reliable software technologies and communication middleware: A perspective and evolution directions for cyber-physical system, mobility, and cloud computing.} 
Future Generation Computer Systems, vol. 71, pp. 171--176. June 2017.

\bibitem{ACMSigbediLand}
M. Garc\'{i}a Valls, N. Herrasti, C. Jouvray, A. Armentia.
\textit{Flexible and timely on-line integration of medical services using iLand middleware}. ACM Sigbed Review. 2017.

\bibitem{gpcRaspi}
M. Garc\'{i}a-Valls, J. Ampuero-Calleja, Luis Lino Ferreira. 
\textit{Integration of Data Distribution Service and Raspberry Pi}. 
Proc. of $12^{th}$ International Conference on Green, Pervasive and Cloud Computing (GPC). LNCS vol. 10232. May 2017. 

\bibitem{gpcUMA}
M. Garc\'{i}a-Valls, D. Garrigo, M. D\'{i}az. 
\textit{Impact of middleware design on the communication performance}. 
Proc. of $12^{th}$ International Conference on Green, Pervasive and Cloud Computing (GPC). LNCS vol. 10232. May 2017. 

\bibitem{DDSPartitioned}
M. Garc\'{i}a-Valls, J. Dom\'{i}nguez-Poblete, I. E. Touahria.
\textit{Using DDS in distributed partitioned systems.}
ACM Sigbed Review. 2017. (To appear)

\end{thebibliography}
\end{document}